# High Space-bandwidth Product Label-free Examination of iPSC-derived Brain Organoids via Fourier Ptychographic Microscopy

Mikolaj Krysa, Mikolaj Rogalski, Piotr Arcab, Pawel Goclowski, Kamil Kalinowski, Piotr Zdańkowski, Vishesh K. Dubey, Mukesh Varshney, Balpreet S. Ahluwalia, Maciej Trusiak

***Abstract*—Fourier ptychographic microscopy (FPM) is a promising quantitative phase imaging technique that enables high-resolution, label-free imaging over a large field-of-view. Here, we present the first application of FPM for the quantitative analysis of human brain organoid slices, providing a powerful, cost-effective, and label-free enhancement to the current gold-standard fluorescence microscopy. Brain organoids, prepared as thin (5 µm) slices, were imaged with a custom-built FPM system consisting of a standard light microscope (4x, 0.2 NA objective) and a 7×7 LED array. This configuration achieved a synthetic numerical aperture of 0.54 and a spatial resolution of approximately 488 nm across an area of 2.077 × 3.65 mm. Fluorescence microscopy was used in parallel for neurons, astrocytes, and nuclei labeling, providing rich fluorescence imaging. Moreover, we designed an automated method to merge classical resolution fluorescence images to visualize the whole brain organoid and align it with the numerically increased space-bandwidth product FPM image. The provided alignment method enables rich phase-fluorescence correlative imaging. Based on the segmentation performed on the stitched fluorescence images, we devised a quantitative phase analysis revealing a higher mean optical thickness of the nuclei versus astrocytes and neurons. Notably, nuclei located in neurogenic regions consistently exhibited significantly higher phase values (optical path difference) compared to nuclei elsewhere, suggesting cell-type-specific biophysical signatures. The label-free, quantitative, and high-throughput capabilities of the FPM approach demonstrated here make it a powerful and accessible tool for future structural and functional studies of whole-section brain organoid development and disease modeling studies.**

This work was supported by the National Center for Research and Development (Project No. WPC3/2022/47/INTENCITY/2024 as part of the 3rd competition for joint research projects as part of Polish-Chinese cooperation (2022), and by BSA European Union's HORIZON Research and Innovation Actions under grant agreement No. 101191315. Corresponding authors: Mikolaj Krysa, Maciej Trusiak

Mikolaj Krysa, Mikolaj Rogalski, Piotr Arcab, Kamil Kalinowski, Piotr Zdankowski, and Maciej Trusiak are with the Institute of Micromechanics and Photonics, Warsaw University of Technology, 8 sw. Andrzeja Boboli st., 02-525 Warsaw, Poland (e-mails: mikolaj.krysa@pw.edu.pl, mikolaj.rogalski@pw.edu.pl, piotr.arcab.dokt@pw.edu.pl, kamil.kalinowski.dokt@pw.edu.pl, piotr.zdankowski@pw.edu.pl, maciej.trusiak@pw.edu.pl, respectively)

Pawel Goclowski, Vishesh K. Dubey, and Balpreet S. Ahluwalia are with the Department of Physics and Technology, UiT The Arctic University of Norway, Tromsø 9037, Norway (e-mails: pawel.goclowski@uit.np, vishesh.k.dubey@uit.no, balpreet.singh.ahluwalia@uit.no, respectively)

Balpreet S. Ahluwalia is also with the Department of Physics, University of Oslo, 0313 Oslo, Norway.

Mukesh Varshney is with the Division of Clinical Microbiology, Department of Laboratory Medicine, Karolinska Institutet, Huddinge, 141 52, Sweden (e-mail: mukesh.varshney@ki.se)

## I. INTRODUCTION

BRAIN organoids represent a revolutionary advancement in biomedical research, offering three-dimensional, self-organizing neural tissue cultures derived from human induced Pluripotent Stem Cells (iPSCs) that bridge the gap between traditional cell cultures and animal models [1], [2]. These models recapitulate human-specific features of brain development, including the expanded outer subventricular zone and outer radial glial cells characteristic of primate neurogenesis, which are absent in rodent models. This human relevance is crucial given that approximately 90% of drugs showing promise in animal studies fail in clinical trials due to species-specific differences [3]. Brain organoids excel in disease modeling for neurodegenerative conditions and drug discovery applications, providing physiologically relevant platforms for testing therapeutic compounds. Their compatibility with advanced imaging modalities, combined with label-free analysis capabilities, makes them an ideal candidate for quantitative investigations of brain development and pathology [4].

The current gold standard for the brain organoid examination is fluorescence microscopy. Its ability to image fluorescent tags, which label selected antigens in the tissue (specific organelles or cell types), with the popularity of this imaging modality in biomedical laboratories, makes it a common, undeniably helpful, and sometimes a vital choice [5]. It has, however, some limitations, including issues with photobleaching and photo-toxicity, as well as the inability to image structures other than the fluorescent-labeled ones. These limitations are often tackled by utilizing a label-free optical microscopy method, such as phase-contrast or differential interference contrast microscopy, in the same equipment. Despite these two methods enhancing the contrast of the observed samples by utilizing the difference in refractive index, they suffer from two other limitations: the contrast increase is limited, and typically, these techniques use the same objectives as the confocal microscope, which limits their field of view (FOV) [6]. These limitations might be subdued using Fourier ptychographic microscopy (FPM).

FPM is a computational quantitative phase imaging (QPI) technique. Like many other QPI methods, it enables label-free imaging of the transparent samples by measuring sample-induced phase delay, which linearly correlates with sample thickness and its refractive index [7], [8], [9]. However, unlike other QPI frameworks, FPM does not face a fundamental microscope-objective-driven spatial resolution versus FOV trade-off. It works by capturing several low-resolution images – usually using a low numerical aperture (NA) and a large FOV objective – from different angles and then stitching them together in the Fourier space to produce large FOV, high-resolution complex images [10]. It therefore enables obtaining the amplitude (absorption-based contrast) and phase (refraction-based contrast) images with a FOV of a low NA objective, but with synthetically increased resolution [10], [11].

Typically, samples prepared for fluorescence microscopy might also be imaged using an FPM system [12]. Moreover, the FPM system can be adapted to a light microscope (available in most biological labs) at a very low cost by replacing the light source with an LED array. Additionally, open-source software for the reconstruction of FPM images is available, which makes FPM an ideal choice for the low-cost improvement of imaging systems to study biomedical samples [13].

Several studies have used quantitative phase imaging techniques for organoid investigation. Quantitative oblique back-illumination microscopy (qOBM) was used for brain organoids [14], [15], while phase contrast and holotomography were used for intestinal organoids [16], [17]. However, all these methods face trade-offs between spatial resolution and FOV, which can be overcome by FPM.

FPM has been successfully applied in a range of biomedical contexts, from histopathology, such as imaging stained colon cancer sections [19] and unstained, paraffin-embedded breast and fibroadenoma tissues [20], to cytology, including stained fine-needle aspiration smears [21] and blood smears for parasite detection [22] or cell counting [23]. It has also been used for live cell culture monitoring [24], [25], demonstrating its versatility in both fixed and dynamic biological samples. Despite these advances, FPM has not yet been explored for organoid research. In this study, we extend FPM application to brain organoid imaging, leveraging its high space–bandwidth product to capture large-area, high-resolution, label-free images, thereby opening new avenues for correlative and quantitative organoid analysis. Moreover, we provide a method for automatic fluorescence image stitching based on the FPM template and alignment of the stitched image to the FPM phase image. Furthermore, we derive quantitative phase analysis using fluorescence priors, showcasing FPM's high sensitivity to different regions in brain organoid slices, promoting future potential in label-free cell-specific whole slide imaging.

## II. Materials and Methods

### A. Brain organoid preparation

All procedures involving human iPSCs were conducted in accordance with ethical standards and regulations, with approval from the Swedish Ethical Review Authority (Etikprövningsmyndigheten, Uppsala, Sweden; approval number Dnr 2024-05169-01). Brain organoids were derived from iPSCs based on the method described previously [2]. Briefly, human iPSCs KISCOi001-A were differentiated into cortical organoids. The iPSCs were cultured and collected in a single-cell suspension as described earlier [26]. The single cell suspension was transferred to AggreWell 800 plates (STEMCELL) to generate 3D cellular aggregates/neurospheres in a neural induction medium (FB1) containing dual SMAD inhibitors and 10 μM ROCK inhibitor (Medchem Express) for two days, followed by a fresh media change without ROCK inhibitor with daily changes until day 6. On day 7, the developing brain organoids were embedded in basement membrane extract (Geltrex, Life Technologies) to make domes in a 100 mm Petri dish, placed in the incubator for 20 minutes for gelation, followed by the addition of neural expansion media (FB2) and cultured with daily media changes until day 10. On day 11, the domes were detached and transferred to 12-well ultra-low attachment plates (Corning) using a spinning bioreactor, as described earlier [27]. The domes were cultured in FB2 media until day 24, with media changes every other day. On day 25, FB2 media was replaced with neural differentiation media (FB3), and brain organoids were cultured until day 90 with media changes every 3-4 days. Fig. 1 shows a schematic workflow for brain organoid derivation from human iPSCs. Culture media composition at every developmental stage is shown in Table S1. The 90-day culture period was selected to achieve mature spatial organization with established neuroproliferative zones and radial cell distribution patterns, as characterized by immunofluorescence analysis of developmental markers. The process is shown in Fig. 1.

### B. Immunofluorescence staining and microscopy

Three Brain organoids from a batch, at 90 days of culture, were fixed in 4% paraformaldehyde (Santacruz) overnight at 4°C, followed by 3 washes in PBS and processed in 30% sucrose overnight for cryosectioning. 5μm thick cryosections/slices were cut on Cryostar Nx 70 (ThermoFisher). Sections were immunostained with mature neuron marker MAP2 (ThermoFisher), astrocyte marker GFAP (Abcam), and nuclei were labeled with DAPI. Briefly, sections were washed in PBS for 5 minutes, followed by permeabilization with 0.2% Triton X-100 and normal serum blocking for 30 minutes. Primary antibody cocktail was applied on the section and incubated at 4°C overnight, followed by three washes of 5 minutes each in PBS, and Alexa Fluor labelled secondary antibody cocktail of Donkey anti-Chicken (555) and Donkey anti-Mouse (488) for one hour at room temperature. This was followed by three washes in PBS and incubation with 300 nM DAPI for 10 minutes. Sections were mounted with Pharamount (Dako) mounting media with #1.5 coverglasses (ThermoFisher) and cured for 20 minutes. The imaging was performed on a commercial Delta Vision microscope. Three fluorescence channels were used for this purpose: red for MAP2 (542 nm excitation wavelength, exposure time = 50 ms), green for GFAP (475 nm excitation wavelength, exposure time = 150 ms), and blue for DAPI (390 nm excitation wavelength, exposure time = 400 ms). Low-magnification images of the whole organoids were acquired with a 20×/0.8NA objective lens, while detailed fluorescence images from Fig. 2 were acquired with a 60×/1.42NA objective lens with oil immersion. Rejection of out-of-focus light and image quality improvement were achieved via a deconvolution operation, performed by the microscope manufacturer's software. Due to the limited FOV of the system, multiple images (between 5 and 12) had to be acquired and stitched together afterwards to obtain comprehensive, detailed images of the whole organoid slices.

### *C. Fourier Ptychographic Microscopy*

The FPM system used in this study was a typical brightfield microscope (Nikon Eclipse Ei) with a 4x 0.2 NA objective (Nikon Plan Apo λD) and a 2.74 µm camera pixel size (Basler a2A5320-23umBAS CMOS camera), but without a condenser and with a 7x7 LED array as a light source ~70 mm below the sample. Green light LEDs (centered at 523 nm, with a distance of 8.1 mm apart from each other) were used in this study. All 49 images illuminated from different angles were acquired and used to reconstruct the resolution-increased phase image using the Quasi-Newton algorithm [28] (synthetic NA≈0.54 – experimentally checked; resolution comparison shown in Fig. S1). The images were reconstructed using an open-source MATLAB application [13].

### *D. Fluorescence image preprocessing, alignment, and segmentation*

All the fluorescence microscopy images were normalized to the top 99.99th percentile. Afterwards, the fluorescence images were aligned to the FPM images. To achieve this, the phase image was first preprocessed to resemble the fluorescence image. Specifically, the median value was subtracted from the phase image, negative values were set to zero, and the result was normalized. This produced an image with a background close to zero, visually similar to the fluorescence image. We then applied the KAZE [29] algorithm to both images to extract keypoints. Each detected keypoint was assigned a descriptor - a vector encoding the local neighborhood around the point. Next, keypoints from both images were matched in pairs, so the overall descriptor matching error was minimized. These point pairs implicitly captured the spatial relationship between the two images. To estimate an affine transformation from these correspondences, we used the MSAC (M-estimator SAmple Consensus) algorithm [30]. The resulting transformation matrix was used to align the original-size fluorescence images, which was performed subsequently. In cases where fluorescence images overlapped, pixel values were averaged.

Since KAZE is computationally expensive, we downsampled the phase image by a factor of 4 and the fluorescence image by a factor of 8 before applying the algorithm. This downsampling was later compensated for by adjusting the affine transformation matrix. Assuming the transformation matrix $T'$ was computed using the downsampled images (with scaling factors $s_{\text{ph}} = 1/4$ for phase and $s_{\text{fl}} = 1/8$ for fluorescence), we rescaled it to match the original image resolutions. The linear part of the matrix was scaled by (1)

$$r = s_{fl}/s_{ph} \tag{1}$$

The corrected affine matrix $T$ was obtained as follows:

$$\mathbf{T}' = \begin{bmatrix} a'_{11} & a'_{12} & 0 \\ a'_{21} & a'_{22} & 0 \\ t'_x & t'_y & 1 \end{bmatrix} \quad \Rightarrow \quad \mathbf{T} = \begin{bmatrix} r \cdot a'_{11} & r \cdot a'_{12} & 0 \\ r \cdot a'_{21} & r \cdot a'_{22} & 0 \\ \frac{t'_x}{s_{\text{ph}}} & \frac{t'_y}{s_{\text{ph}}} & 1 \end{bmatrix}$$

Where $a'_{ij}$ are coefficients describing linear transformations such as rotation, scaling, and shear, while $t'_x$ and $t'_y$ represent translations along the x and y axes.

The linear part is scaled by $r$ , and the translation components are divided by $s_{ph}$, yielding $T$ in the coordinate system of the original images.

After the alignment and merging of the fluorescence images, they were segmented using the Otsu algorithm [31]. After the segmentation, all the objects smaller than 150 pixels (4.4376 µm$^2$) were removed from each segmentation mask, due to the high probability of false detection. Nuclei masks were further preprocessed for a more accurate segmentation. This preprocessing included further segmentation of big structures (over 5000 pixels – 147.92 µm$^2$) that were not separated using the Otsu algorithm. The 0.7 quantile was calculated from these structures, and the pixels with values lower than the 0.7 quantile were removed. Afterwards, the objects smaller than 300 pixels were removed, and the rest of the detected structures were smoothed using 3-pixel radius disks. Finally, the holes inside each object were filled. All the above-mentioned analyses were performed in MATLAB (ver. R2024b).

### *E. Descriptive statistics of cell types/structures phase values*

The segmentation masks from each channel were acquired and used to obtain phase values from astrocytes, neurons, and nuclei. The phase values were used to calculate descriptive statistics: mean, median, standard deviation, first and third quartiles (0.25 and 0.75 quantiles, respectively). Moreover, the distributions of the values were evaluated using the Lilliefors test. These statistics were also calculated for the nuclei mask that overlaid with the detected cell types (mature neurons and astrocytes) and for the nuclei that did not overlay with any detected cell types. Apart from the analysis of the whole organoid slices, some of the interesting regions were chosen from each of the slices. All of these analyses were done using MATLAB (ver. R2024b).

## III. Results

The FPM imaging was performed on three organoid slices to evaluate whether the FPM system is capable of visualizing the cellular composition of brain organoid slices. The 5 µm thick brain organoids were almost invisible in the amplitude image (Fig. S2). However, the phase imaging enabled clear visualization of the brain organoid morphology, see Figs. 3A, 3C, 4A, 4C, and S3A. Moreover, all the slices could be visualized in a single image with a sub-cellular resolution (experimentally tested spatial resolution of the used setup was ~488 nm).

The fluorescence images were overlaid on the FPM phase images to evaluate how the brain organoid slices' phase morphology aligns with cell types/structures (Figs. 3B, 3D, 4B, 4D, and S3B). The whole organoid fluorescence image was preprocessed and merged before overlaying (described in the Materials and Methods section). The structures stained fluorescently were: soma and dendrites of mature neurons (MAP2+), astrocyte cytoplasm (GFAP+), and the cell nuclei (DAPI staining). Unfortunately, the astrocytes were absent in slice number 3 (Fig. S3). Nonetheless, the results show a good visual alignment of the investigated cell types/structures with the FPM phase images (Fig. 3, 4, and S3).

To further validate the cross-modality alignment, the root mean square error (RMSE) between automatically detected landmark points in the fluorescence and FPM images was computed after applying the estimated translational shift. It yielded residual errors of approximately 1.4 pixels across representative samples. Although this demonstrates strong internal consistency of the shift-estimation algorithm, these landmarks originate from the same automatic feature-extraction pipeline and therefore cannot serve as an independent measure of true alignment accuracy. To overcome this limitation, manual annotation of distinct, clearly identifiable structures observable in both modalities was also performed. Depending on local morphology, these manually selected features included centers of circular objects, pronounced edges, or isolated bright spots (Fig. S4). The resulting RMSE values ranged from 1.884 to 3.563 pixels, with larger discrepancies observed in regions exhibiting greater structural heterogeneity. Taken together, the automatic and manual analyses indicate that the applied alignment procedure is robust and reproducible across samples. The manual evaluation, despite involving fewer landmarks, independently confirms that the residual inter-modality misalignment remains small (0.324-0.613 µm) relative to the relevant structural features (35-251 µm²), thereby supporting the reliability of subsequent cross-modality comparisons.

Afterwards, the phase values from mature neurons, astrocytes, and nuclei were obtained to further evaluate the structures' correspondence with a phase image. Moreover, the phase values of nuclei colocalized with one of the cell types and the values of nuclei that were not colocalized with any of the structures were acquired. The values were used to calculate descriptive statistics (Table 1).

Since astrocytes were not present in slice number three, all of the results showing astrocyte results in Table 1 are the mean of slices one and two. All of the above-mentioned phase values had a distribution other than normal (shown using Lilliefors' test). The median and interquartile range of mature neurons and astrocytes were quite similar, while they were much higher in the nuclei. Moreover, the median phase values were even higher when nuclei were overlaid with any other investigated structure. Interestingly, the median values of the nuclei that did not overlay with any investigated structures were slightly lower than all of the nuclei due to the exclusion of some of the higher value (overlapping) nuclei.

Phase value distributions (medians and interquartile ranges) remained highly consistent across slices for most structures, as evidenced by histograms (Fig. 5A–C) and quantitative data (Tables 2, 3, S2). Astrocytes were the exception, showing slight variations between slices 1 and 2. Consistent with the three-slice mean: a) Nuclei exhibited higher phase values than mature neurons and astrocytes; b) Overlaid structures showed elevated phase values compared to non-overlaid regions; c) Nuclei without structural overlays displayed the lower phase values in all slices. The difference between phase values of the nuclei and immunostained structures might be due to different binding sites of the stains/antibodies. DAPI, which stains nuclei, binds to DNA [32], while the astrocyte and neuron markers (GFAP and MAP2, respectively) are proteins that build the cytoskeleton and microtubules (respectively), and reside in the cytoplasm [33], [34]. Therefore, the quantitative phase analysis revealed biologically meaningful differences between cellular compartments. Nuclei consistently showed higher phase values (mean: 0.253 radians) compared to neuronal (0.151) and astrocytic (0.156) cytoplasm, reflecting the higher refractive index of condensed chromatin and nuclear proteins, which is consistent with a study of nuclei' refractive index in 3D structures [35]. The similar phase values between neurons and astrocytes suggest comparable cytoplasmic densities in their soma regions, consistent with their shared glial-neuronal lineage and similar organelle distributions. These phase measurements provide quantitative metrics for cellular density and organization that complement traditional fluorescence-based approaches.

The phase values of nuclei in neurogenic regions were obtained to investigate whether these cells have an increased optical density of nuclei compared to other brain organoid cells. The regions highlighted in Fig. 6 demonstrate characteristics consistent with active neurogenic zones in brain organoids [36]. These areas exhibit the typical morphological features, including high cellular density and organized radial architecture of ventricular-like structures, where neural stem cells and progenitor cells are actively dividing. The descriptive statistics from these regions were calculated and compared to the same statistics from the whole section. The median phase value of the neurogenerative cell nuclei was ~24% higher in slice one (0.277 compared to 0.224 in the entire slice), ~14% higher in slice two (0.256 compared to 0.224 in the whole slice), and ~16% higher in slice three (0.302 compared to 0.261 in the entire slice). The elevated phase values in these regions reflect increased cellular density due to proliferating neural stem cells and intermediate progenitor cells, which require additional cellular components for active cell division. These components include elevated ribosome content, an expanded endoplasmic reticulum, and condensed chromatin during the S and G2 phases of the cell cycle [37].

To evaluate whether the FPM system presented in this study is sufficiently quantitative to draw the above-mentioned conclusions, its accuracy was validated with a quantitative phase target (Benchmark Technologies Quantitative Phase Target (QPT™)). The large structures in the target displayed a gradual phase decrease toward their centers, consistent with the theory of FPM reconstruction [38] (Fig. S5A). However, considering the features of the size analyzed in this study (neuron fragments, astrocyte fragments, and nuclei), which typically ranged from 35 μm² to 251 μm², they showed only a minor variation in phase values (0.296–0.338 rad) (Fig. S5A, B). Furthermore, for structures of this size, the phase values increased linearly with the thickness of the calibration samples: 50 nm structures showed an average phase value of 0.18 rad, 100 nm structures averaged 0.32 rad, and 150 nm structures averaged 0.40 rad (Fig. S5C, D). These results indicate that, while our system is not fully quantitative in absolute phase values, its performance is sufficient for comparative analysis in relative phase values across the object size range studied – from the smallest neuronal fragments to the largest connected nuclear populations.

FPM imaging successfully visualized the brain organoid architecture across three slices, revealing cellular organization patterns consistent with those of human cortical development. The 5 μm slice thickness preserved cellular morphology while enabling phase contrast detection of density variations between different neural cell populations. Phase imaging revealed distinct organizational zones reminiscent of ventricular and subventricular regions found in the developing human cortex.

## IV. Discussion

This study demonstrated that FPM enables imaging of whole-brain organoid sections with subcellular spatial resolution. Using a 4x objective lens with NA of 0.2, we could image an area of approximately 7.58 mm² (2.077 x 3.65 mm), and synthetically increase the NA to 0.54 (increase the spatial resolution from 1380 nm to 488 nm; the resolution increase is shown in Fig. S1). This large FOV and high resolution combination distinguishes FPM from other phase imaging techniques.

The critical step for the success of merging fluorescence images and aligning merged images to FPM phase images was image processing. It was a challenge since the fluorescence images had different spatial and pixel sizes and were usually slightly rotated compared to the FPM image. Before the alignment, the fluorescence images had to be normalized to the top 99.99th percentile since some pixels with extremely high values mask the rest of the immunostained places. Afterwards, phase images were preprocessed by subtracting the median and setting negative values to zero, making them visually more similar to fluorescence images. Fluorescence channels were summed and contrast-adjusted. Both image types were downsampled to enable robust keypoint matching, and the KAZE algorithm was used for feature detection. It outperformed all other tested detectors, including Harris [39], MinEigen [40], FAST [41], SURF [42], BRISK [43], and ORB [44], particularly when combined with downscaling. Affine transformations estimated from matched keypoints were rescaled and applied to align the fluorescence images. Alternative matching methods, such as DFT-based registration [45] and cross-correlation [46], were also evaluated but failed due to scale and image characteristics differences, even after preprocessing. This feature-based, fully automated approach enables accurate large-area fluorescence reconstruction using the FPM image as a common reference, minimizing manual effort and improving throughput, especially in cases requiring many acquisitions like organoid slices.

Previous phase imaging studies examining organoids have generally focused on live, unsliced specimens. For example, some researchers used qOBM to image brain organoids, but the imaged area was smaller than in our study [14], [15]. Other groups applied phase contrast or holotomography to intestinal organoids, but using higher magnification objectives resulted in a smaller FOV compared to our approach [16], [17]. One study on retinal organoids employed OCT and micro-coherence tomography [18]. While optical coherence tomography offered lower spatial resolution than our method, micro-coherence tomography achieved a similar resolution (1–2 μm versus our ~0.49 μm). Still, it covered a smaller area (about 1.25 x 1.25 mm versus our 2.077 x 3.65 mm), making it comparable but still less effective overall. Additionally, micro-coherence tomography systems are significantly more expensive than the simple light microscope and LED array required for FPM.

Our FPM setup is highly cost-effective – it consists of a standard light microscope with a 4x, 0.2 NA objective, a camera, and a 7x7 LED array. Since most biological laboratories already possess a light microscope, only minimal modifications are needed. Implementation is straightforward, involving the replacement of the light source and condenser with an LED array. Moreover, automatic methods for LED array calibration are available (since the central LED must be precisely aligned with the optical axis), which further simplifies the implementation [47]. However, there is no open-source software for automated data acquisition (it needs to be tailored to an LED array), but the open-source tools for data reconstruction are available. One of them is an “FPM app” that provides a user-friendly interface, enabling the use of different FPM setups (including different LED array sizes, collection sequence, distance from the sample, and objective parameters), and making FPM image reconstruction significantly easier [13].

The phase value differences presented in this study reflect fundamental cellular properties. Higher phase values in nuclei compared to cytoplasm result from the dense chromatin organization and nuclear protein content. The elevated values in neurogenic regions indicate active cell cycle progression, where proliferating cells accumulate cellular components necessary for division. This correlation between phase measurements and cellular activity demonstrates FPM's potential for label-free monitoring of neurogenesis dynamics.

Future applications include disease modeling, where altered patterns of neurogenesis characterize pathological conditions. Autism spectrum disorders, microcephaly, and neurodegenerative diseases exhibit distinct neurogenic abnormalities that could be quantitatively assessed using phase-based measurements. Additionally, drug screening applications could utilize phase value changes as endpoints for compounds affecting neural stem cell proliferation or differentiation. Moreover, the label-free nature of FPM analysis could enable the investigation of living organoids, which may facilitate real-time studies of developmental timing, drug responses, and disease progression in the future. However, studying unsliced brain organoids would require significant improvements in setup. For example, the use of tomographic techniques, such as 3D FPM [48], [49] or intensity diffraction tomography [50], [51], to distinguish cells at different depths would be required. Moreover, intact brain organoids are significantly thicker than the slices used in this study, which would cause multiple scattering and hinder the use of FPM. This would require the use of reconstruction methods taking into account multiple scattering [52], [53] or using methods to numerically [54] or physically [55] reduce scattering of the samples (e.g., optically clear the samples). Additionally, the study presented here employed fluorescence-guided, label-free quantification of cell types and structures in the brain organoids. However, to fully leverage the label-free nature of FPM and introduce only the environmental factors relevant to the study topic, immunostaining should not be employed. Nevertheless, the knowledge of cell types is often required to answer key biological questions. While the mere phase values did not allow for full distinction of the cell types and nuclei for now, they might be used in the future as one of the properties, along with cell shape, in conjunction with computer vision and machine learning methods (similarly to virtual staining [17], [56]), to distinguish them in a future fully label-free investigation [57]. Moreover, to study dynamic processes, the acquisition time should be as short as possible. In our case, the acquisition took approximately 2.5 minutes, and the image reconstruction took about 1.5 minutes, which is perfectly suited for imaging static slices, but is definitely too long for a study of dynamic processes, such as acute drug responses [58]. To enable it, the acquisition times should be optimized. Therefore, higher-speed, highly sensitive cameras and brighter LED arrays should be utilized. Other researchers have shown that, with proper optimization, FPM imaging can be performed even with a single shot, thereby reducing the time to the speed limit of camera acquisition [59], [60], [61]. If all of the above-mentioned improvements were met, fully label-free studies of living organoids could be performed.

## V. Conclusions

Our results demonstrate that a low-cost FPM system can image entire brain organoid slices (up to 2.077 x 3.65 mm) with a resolution of ~488 nm. We also demonstrated that automated feature detection and affine transformation enable efficient alignment and merging of multiple fluorescence images with FPM phase images. Phase imaging provides additional information about cell types and structures that antibodies may not label (such as axons not stained by MAP2). It can reveal structures not easily visualized by fluorescence alone. Moreover, the quantitative nature of the FPM system enables the extraction of additional information from the samples, revealing biologically meaningful phase signatures that correlate with cellular identity and neurogenic activity. The ability to detect neurogenic regions without fluorescent labeling represents a significant advancement for developmental neurobiology and disease modeling, which could be utilized to deepen the insight from future organoid studies.

Figures and tables

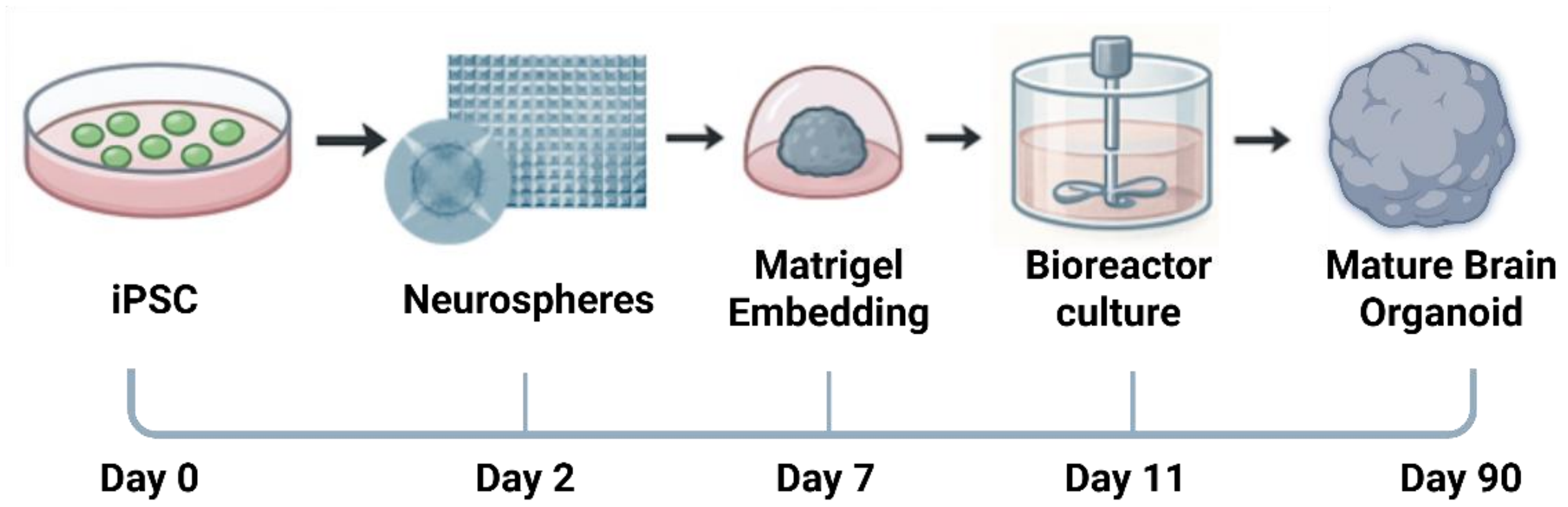

**Fig. 1.** Brain organoid derivation from human iPSCs. Protocol for brain organoid derivation. Human iPSCs are differentiated into neurospheres using AggreWell plates with neural induction medium containing dual SMAD inhibitors (Days 0-2). At Day 7, neurospheres are embedded in Matrigel domes and cultured in neural expansion medium. By Day 11, organoids are transferred to a spinning bioreactor culture for enhanced nutrient distribution and maintained in neural differentiation medium until Day 90, when mature brain organoids with complex three-dimensional architecture are achieved.

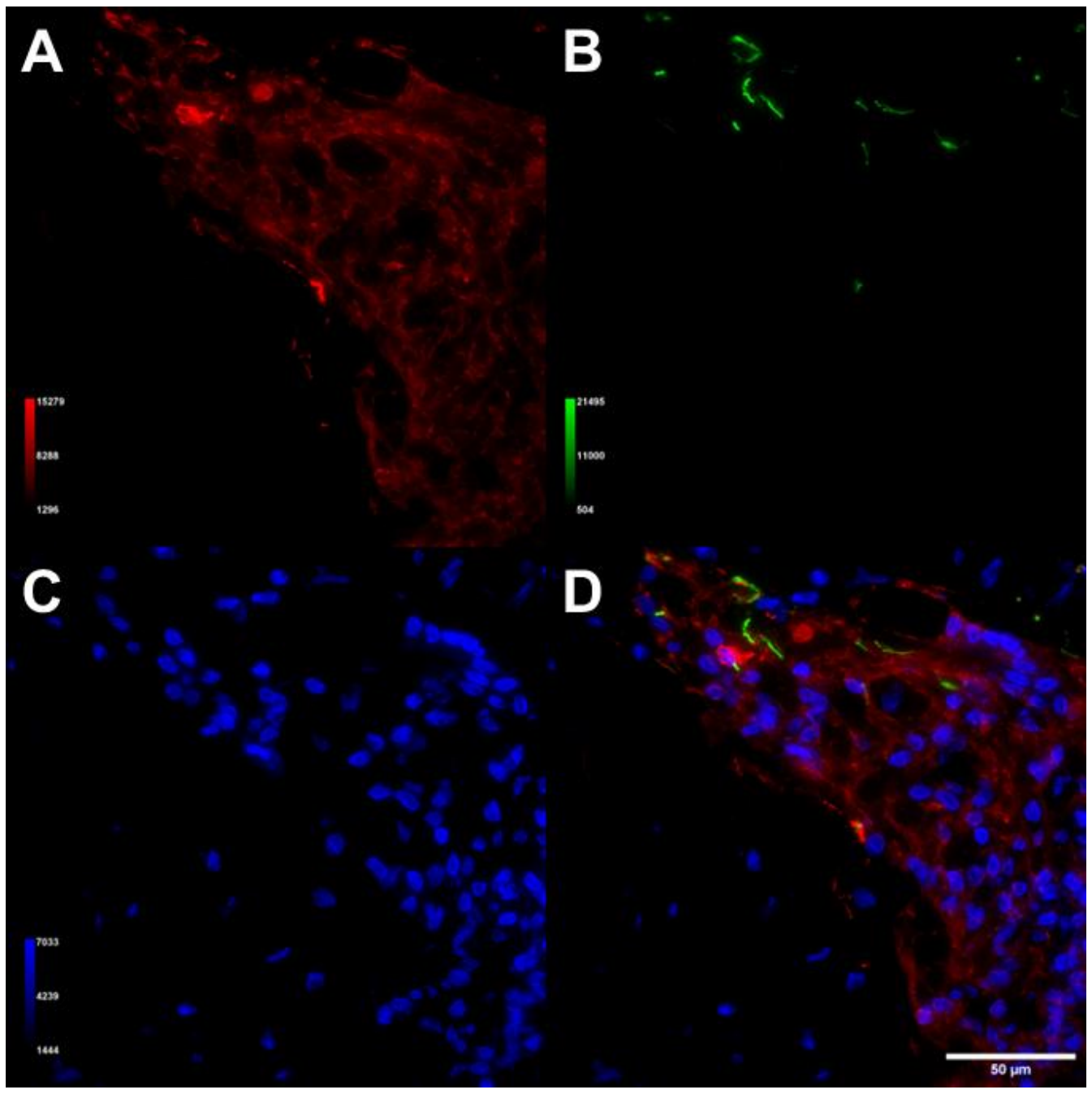

**Fig. 2.** Fluorescence imaging of a selected region of the brain organoid. Immunofluorescently stained somas and dendrites of differentiated neurons (red, MAP2 marker) (A), immunofluorescently stained cytoplasm of astrocytes (green, GFAP marker) (B), DAPI-stained cell nuclei (blue) (C), and the overlay of all the channels. The imaging was performed using 60×/1.42NA objective lens with oil immersion.

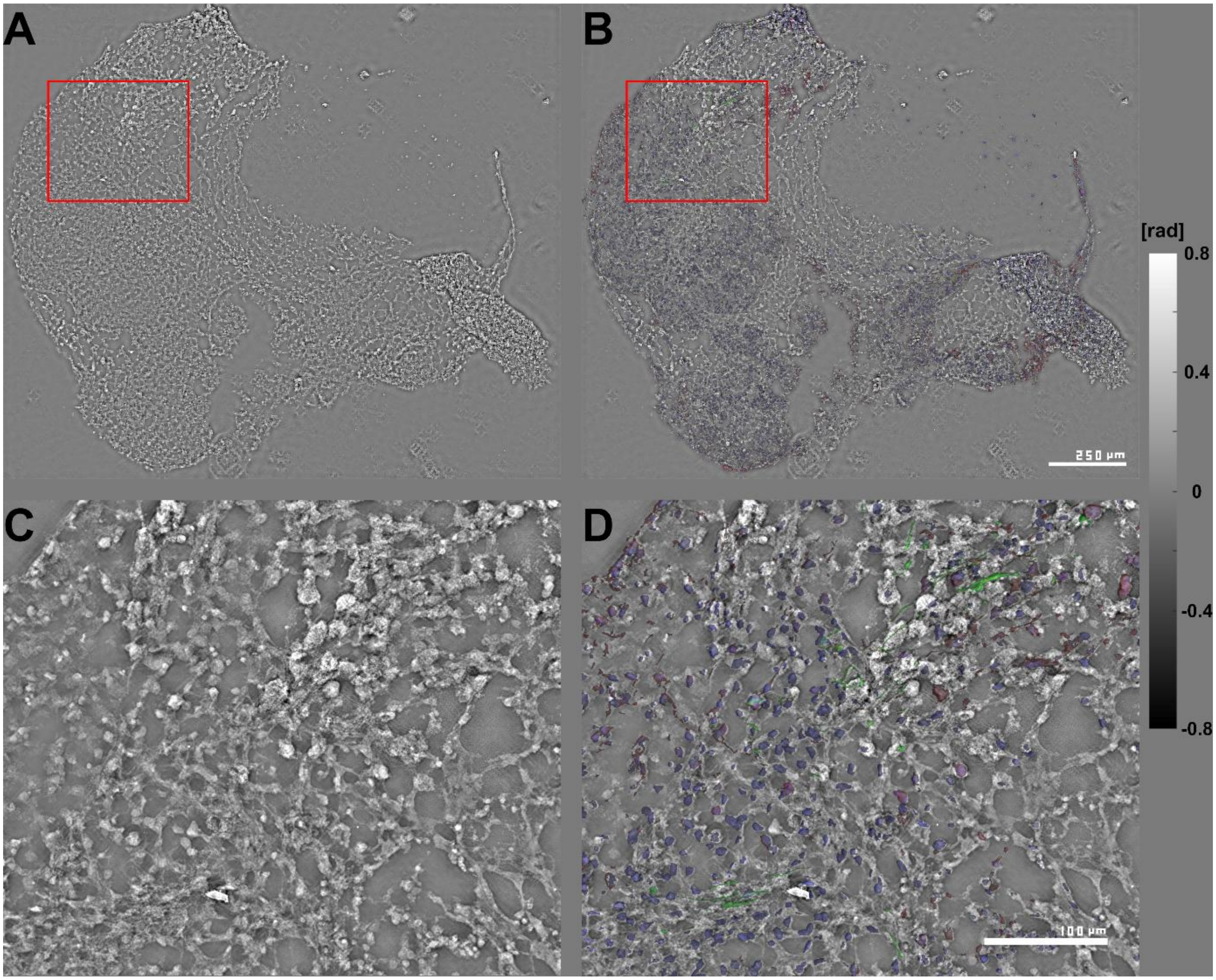

**Fig. 3.** FPM phase representation (A, C), fluorescence microscopic image overlayed on FPM phase image (B, D), of the whole slice (A, B), and a zoom to a selected region (C, D) of the first slice of the brain organoid. The red color represents mature neurons (MAP2 immunostaining), the green color represents astrocytes (GFAP immunostaining), and the blue color represents cell nuclei (DAPI staining). The red rectangle on the full-size image represents the zoomed area. The color bar on the side represents the phase values (in radians).

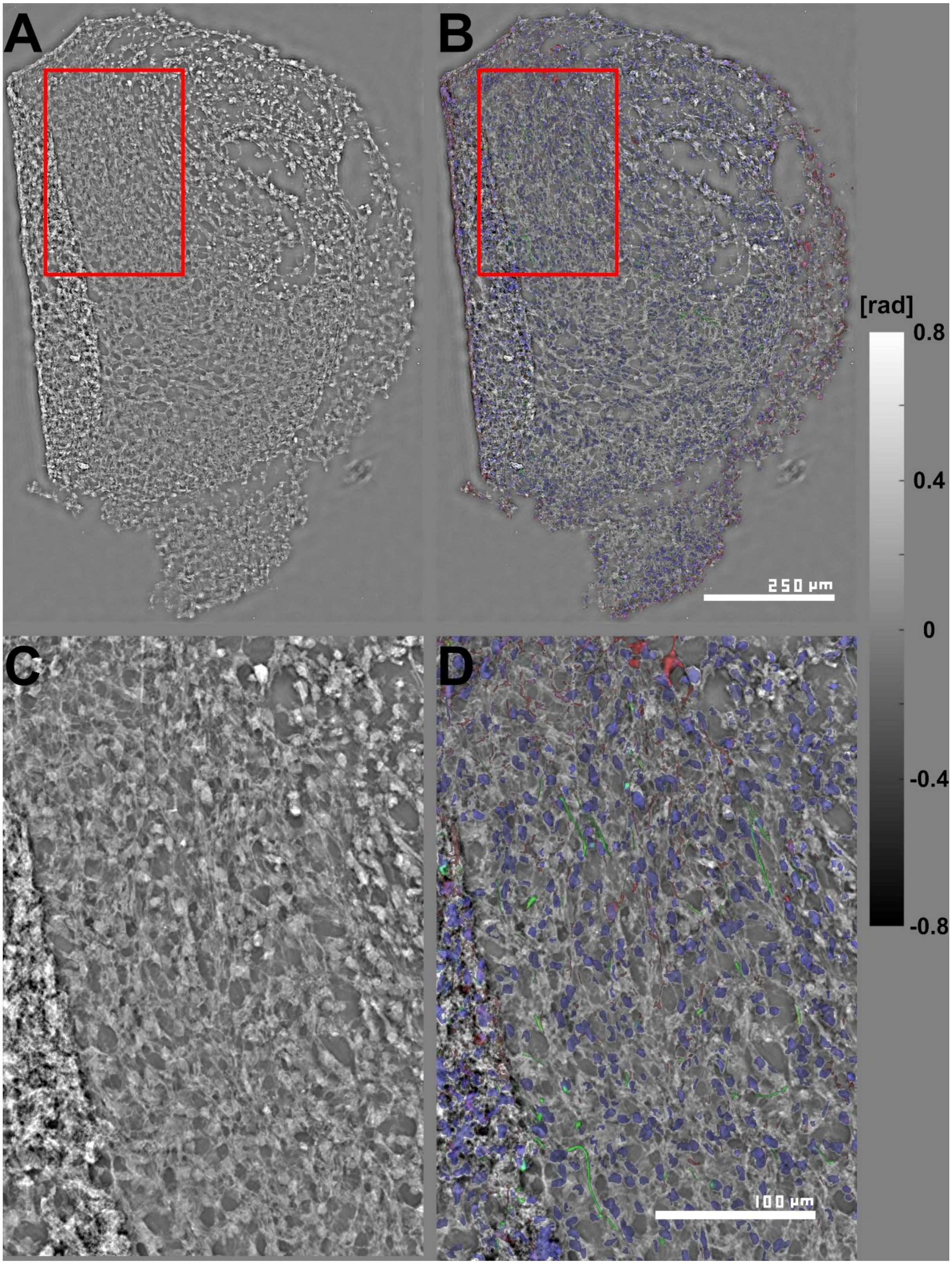

**Fig. 4.** FPM phase representation (A, C), fluorescence microscopic image overlayed on FPM phase image (B, D), of the whole slice (A, B), and a zoom to a selected region (C, D) of the second slice of the brain organoid. The red color represents mature neurons (MAP2 immunostaining), the green color represents astrocytes (GFAP immunostaining), and the blue color represents cell nuclei (DAPI staining). The red rectangle on the full-size image represents the zoomed area. The color bar on the side represents the phase values (in radians).

TABLE I

PHASE VALUE PARAMETERS FROM THE REGIONS CORRESPONDING TO CELL TYPE/STRUCTURES. THE VALUES ARE THE MEANS FROM THREE SLICES. "NEURONS + NUCLEI" AND "ASTROCYTES + NUCLEI" DESCRIBE THE PHASE VALUES OF COLOCALIZED STRUCTURES. "NUCLEI - STRUCTURES" DESCRIBE THE PHASE VALUES OF NUCLEI THAT WERE NOT COLOCALIZED WITH ANY OF THE INVESTIGATED CELL TYPES. "NUCLEI" REFERS TO PHASE VALUES FOR ALL NUCLEI DETECTED.

| | **Mean** | **Median** | **Standard deviation** | **0.25 quantile** | **0.75 quantile** |
|---|---|---|---|---|---|
| **Mature neurons** | 0.151 | 0.134 | 0.237 | -0.004 | 0.291 |
| **Astrocytes** | 0.156 | 0.143 | 0.246 | -0.010 | 0.309 |
| **Nuclei** | 0.253 | 0.236 | 0.224 | 0.105 | 0.386 |
| **Neurons + nuclei** | 0.322 | 0.308 | 0.232 | 0.171 | 0.460 |
| **Astrocytes + nuclei** | 0.348 | 0.343 | 0.239 | 0.192 | 0.499 |
| **Nuclei - structures** | 0.241 | 0.225 | 0.219 | 0.096 | 0.371 |

TABLE 2

PHASE VALUE PARAMETERS FROM THE REGIONS CORRESPONDING TO CELL TYPE/STRUCTURES IN SLICE ONE. NEURONS + NUCLEI AND ASTROCYTES + NUCLEI DESCRIBE THE PHASE VALUES OF COLOCALIZED STRUCTURES. NUCLEI - STRUCTURES DESCRIBE THE PHASE VALUES OF NUCLEI THAT WERE NOT COLOCALIZED WITH ANY OF THE INVESTIGATED STRUCTURES.

| | **Mean** | **Median** | **Standard deviation** | **0.25 quantile** | **0.75 quantile** |
|---|---|---|---|---|---|
| **Mature neurons** | 0.141 | 0.124 | 0.238 | -0.013 | 0.280 |
| **Astrocytes** | 0.127 | 0.111 | 0.258 | -0.048 | 0.292 |
| **Nuclei** | 0.235 | 0.224 | 0.232 | 0.085 | 0.377 |
| **Neurons + nuclei** | 0.306 | 0.300 | 0.245 | 0.152 | 0.456 |
| **Astrocytes + nuclei** | 0.315 | 0.323 | 0.261 | 0.144 | 0.488 |
| **Nuclei - structures** | 0.224 | 0.213 | 0.228 | 0.077 | 0.363 |

TABLE 3

PHASE VALUE PARAMETERS FROM THE REGIONS CORRESPONDING TO CELL TYPE/STRUCTURES IN SLICE TWO. NEURONS + NUCLEI AND ASTROCYTES + NUCLEI DESCRIBE THE PHASE VALUES OF COLOCALIZED STRUCTURES. NUCLEI - STRUCTURES DESCRIBE THE PHASE VALUES OF NUCLEI THAT WERE NOT COLOCALIZED WITH ANY OF THE INVESTIGATED STRUCTURES.

| | **Mean** | **Median** | **Standard deviation** | **0.25 quantile** | **0.75 quantile** |
|---|---|---|---|---|---|
| **Mature neurons** | 0.165 | 0.146 | 0.244 | 0.006 | 0.307 |
| **Astrocytes** | 0.185 | 0.175 | 0.235 | 0.028 | 0.326 |
| **Nuclei** | 0.250 | 0.224 | 0.218 | 0.102 | 0.374 |
| **Neurons + nuclei** | 0.352 | 0.331 | 0.224 | 0.200 | 0.482 |
| **Astrocytes + nuclei** | 0.380 | 0.363 | 0.217 | 0.240 | 0.509 |
| **Nuclei - structures** | 0.237 | 0.212 | 0.214 | 0.094 | 0.358 |

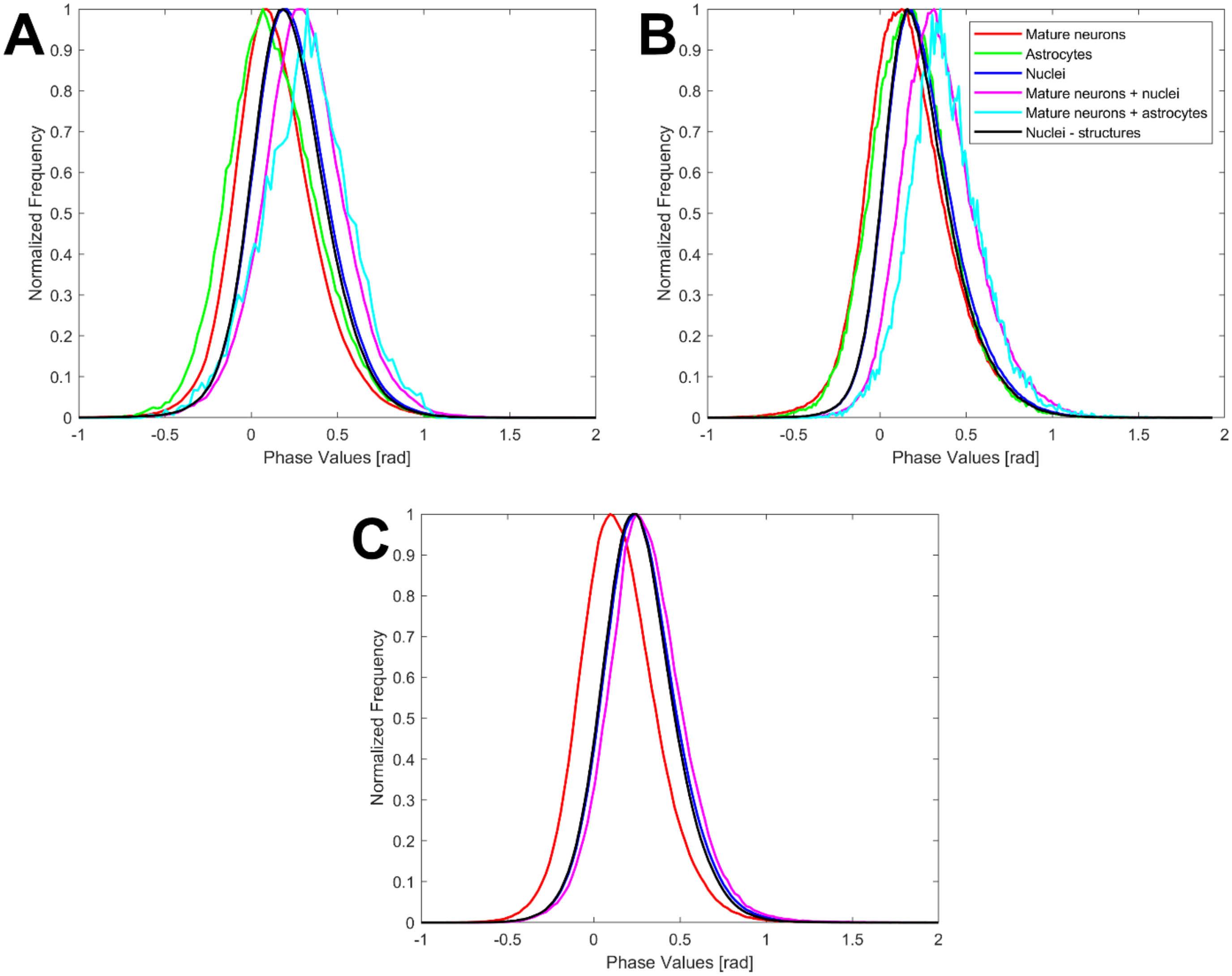

**Fig. 5.** Histograms showing the frequencies of the phase values (in radians) of the FPM phase pixels of different cell types/structures in slices 1-3 (A-C, respectively). Neurons + nuclei and astrocytes + nuclei describe the phase values of colocalized structures. Nuclei - structures describe the phase values of nuclei that were not colocalized with any of the investigated structures.

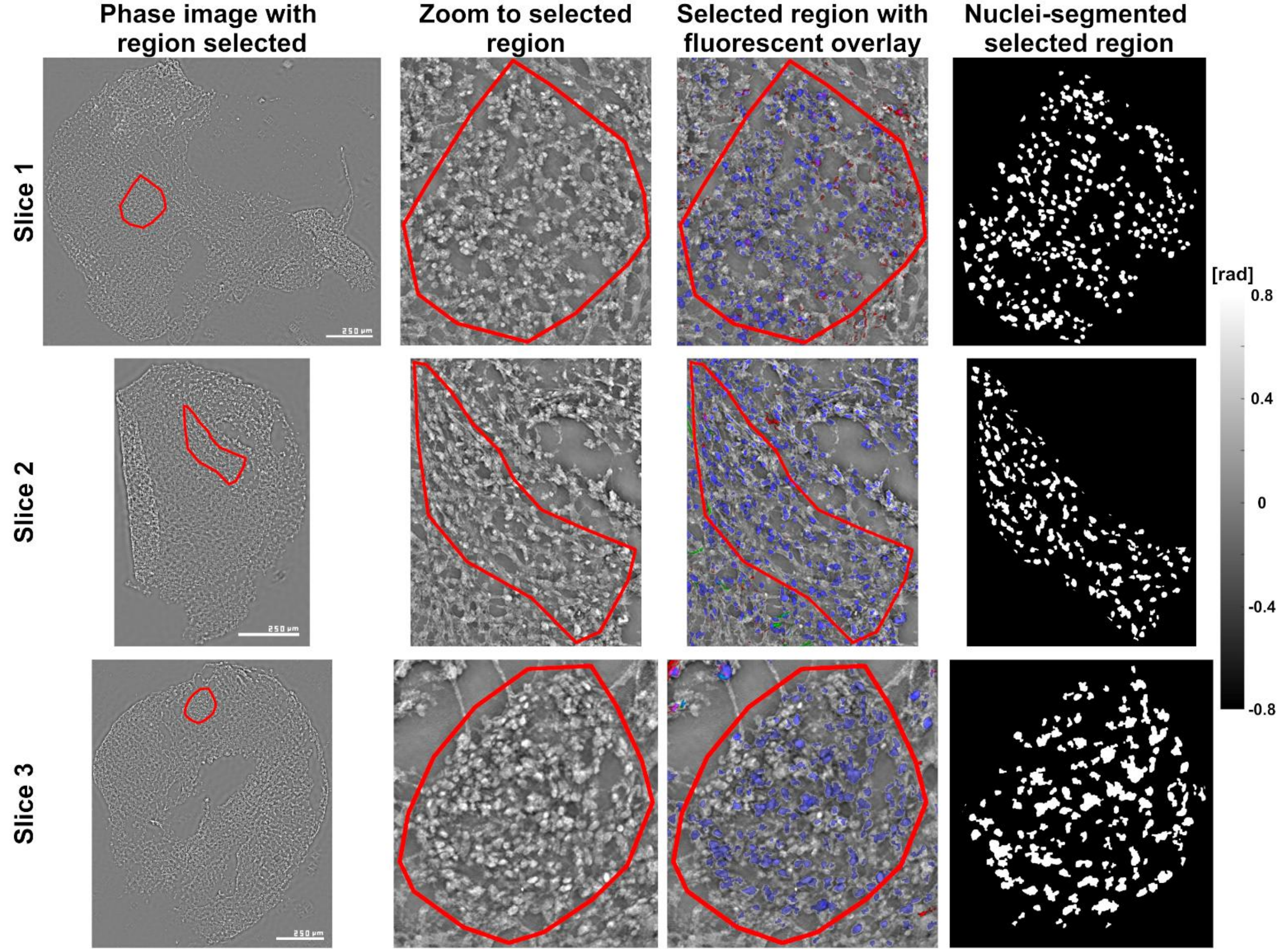

**Fig. 6.** Regions containing generative neurons (regions where neurogenesis occurs) from three brain organoid slices. The regions are framed with a red line. The phase image is a result of the FPM imaging. The blue color on the phase-fluorescence image overlay corresponds to DAPI-stained nuclei. The white spots on the nuclei-segmented selected regions show segmented nuclei, and therefore, the spots where the phase values for the descriptive statistics were obtained.